\documentclass{aastex}
\usepackage{emulateapj5}

\newcommand{\CO}{$^{12}$CO}
\newcommand{\COe}{$CO(1-0)$}
\newcommand{\COz}{$CO(2-1)$}
\newcommand{\vol}[1]{1}
\newcommand{\kms}{km\,s^{-1}}
\newcommand{\cms}{cm^{-2}}

\newcommand{\solar}{L_{\odot}}

\newcommand{\solm}{M_{\odot}}

\slugcomment{}

\shorttitle{CO in the Polar Ring of NGC~2685}
\shortauthors{Schinnerer \& Scoville}

\begin{document}

\title{First Interferometric Observations of Molecular Gas in a Polar Ring:
The Helix Galaxy NGC~2685}

\author{Eva Schinnerer\altaffilmark{1} and Nick Scoville\altaffilmark{2}}
\affil{Radio Astronomy, MS 105-24, California Institute of Technology,
       Pasadena, CA 91125-2400}

\altaffiltext{1}{es@astro.caltech.edu}
\altaffiltext{2}{nzs@astro.caltech.edu}

\begin{abstract}
We have detected four Giant Molecular cloud Associations (GMAs) (sizes
$\leq 6.6'' \approx 430\,pc$) in the western and eastern region of the
polar ring in NGC~2685 (the Helix galaxy) using the Owens Valley Radio
Observatory (OVRO) millimeter interferometer. Emission from molecular
gas is found close to the brightest H$\alpha$ and HI peaks in the
polar ring and is confirmed by new IRAM 30m single dish
observations. The CO and HI line velocities are very similar,
providing additional kinematic confirmation that the CO emission
emerges from the polar ring. For the first time, the total molecular
mass within a polar ring is determined
($M_{H_2}\sim(8-11)\times10^6\,\solm$, using the standard Galactic
conversion factor). We detect about $M_{H_2}\sim4.4\times10^6\,\solm$
in the nuclear region with the single dish. Our upper limit derived
from the interferometric data is lower
($M_{H_2}\le0.7\times10^6\,\solm$) suggesting that the molecular gas
is distributed in an extended ($\ge 1.3\,kpc$) diffuse disk. These new
values are an order of magnitude lower than in previous reports. The
total amount of molecular gas and the atomic gas content of the polar
ring are consistent with formation due to accretion of a small
gas-rich object, such as a dwarf irregular. The properties of the
NGC~2685 system suggest that the polar ring and the host galaxy have
been in a stable configuration for a considerable time (few Gyr). The
second (outer) HI ring within the disk of NGC~2685 is very likely at
the outer Lindblad resonance (OLR) of the $\sim 11\,kpc$ long stellar
bar.

\end{abstract}


\section{INTRODUCTION}

Polar ring galaxies (PRGs) represent an unusual, rare class of objects
which show clear signs of galaxy interaction (Schweizer, Whitmore \&
Rubin 1983). Typically, an early-type (S0 or E) host galaxy is
surrounded by a luminous ring (containing stars, gas and dust) of
$\sim$ 5 to 25~kpc diameter oriented almost perpendicular to the main
stellar disk and rotating about the center of the main stellar body
(see PRG atlas by Whitmore et al. 1990). To-date only about a dozen
PRGs have been kinematically confirmed (e.g. Table 1 in Sparke \& Cox
2000). In the generally accepted picture, the formation of polar
rings is the result of a ''secondary event'': e.g., capture of a
satellite galaxy or accretion of material between (tidally)
interacting galaxies involving a pre-existing S0 galaxy (e.g. Toomre
\& Toomre 1972, Reshetnikov \& Sotnikova 1997). Recently Bekki (1997,
1998) suggested a pole-on merger between two disk galaxies as an
alternative formation mechanism. Observations suggest that polar
rings are long-lived structures (Whitmore et al. 1987, Eskridge \&
Pogge 1997). Possible stabilizing mechanisms are self-gravitation in
the ring (Sparke 1986), or a massive triaxial halo (Whitmore et
al. 1987, Reshetnikov \& Combes 1994)

The Helix galaxy, NGC~2685 ($D\,\sim\,13.5\,Mpc; 1''\,\sim\,65\,pc$;
(R)SB0+ pec) is one of the kinematically confirmed PRGs. Two rings are
detected in HI line emission which have orthogonal angular momentum
vectors (Shane 1980).  Optical and NIR surface photometry (Peletier \&
Christodoulou 1993) suggests an age of 2-6 Gyr for the inner 'polar'
ring and therefore a long-lived structure. The younger HII regions in
the polar ring of NGC~2685 have solar abundances, making accretion of
metal-poor material unlikely (Eskridge \& Pogge 1997).

\section{OBSERVATIONS}

NGC~2685 was observed in its \COe\ line with three pointings covering
the entire polar ring between 2000~April and June using the
six-element Owens Valley Radio Observatory (OVRO) millimeter
interferometer in its C and L configurations. The resulting baselines
(15 - 115~m) provide a spatial resolution of $\sim 6.6''$ (430~pc)
with natural weighting.  The noise per 10\,$\kms$\ channel is $\sim
16$\,mJy\,beam$^{-1}$ in the combined data of 6 tracks. For the
intensity map only emission above the clipping level of $2.5\sigma$
was added together.

The IRAM 30m telescope was pointed towards six positions in NGC~2685
based on the OVRO data (indicated in Fig. \ref{fig:co}). Position N
and W1 were observed on 2002 March 11 (the remaining ones on May 25)
in the \COe\, and \COz\, line using the two 3mm receivers (HPBW $\sim$
21'') and one 1mm receiver (HPBW $\sim$ 11''). The observations had a
total on-source integration time of about 20~min (W2) to 65~min (N)
with an average of 35~min per position. The 1~MHz filter banks
provided a velocity resolution of $\sim 2.6\,\kms$\ ($\sim 1.6\,\kms$)
per channel for the
\COe\, (\COz) line.  The final r.m.s. in a smoothed 10\,$\kms$\ wide
channel was about 4\,mK at 115\,GHz and 6\,mK at 230\,GHz.

We also used archival 21$cm$ HI Very Large Array (VLA) data which are
described in Mahon (1992). The spatial resolution of the combined data
(BnC and D configurations) is 34.6'' $\times$ 33.7'' (12.7'' $\times$
10.8'') for natural (uniform) weighting with a channel width of
20.7\,$\kms$ and an r.m.s.  of $1\sigma\,\approx\,0.3 (0.5)$
mJy\,beam$^{-1}$ for natural (uniform) weighting.

\section{DISTRIBUTION AND KINEMATICS OF THE ATOMIC AND MOLECULAR GAS}

\subsection{Atomic Gas}

The atomic gas forms two distinct rings with radii of $\sim$ 2.4' and
$\sim$ 0.57' which can be easily kinematically distinguished in the
VLA channel maps (Fig. 3.3 of Mahon 1992). We calculated separate data
cubes which contain only emission from the inner and outer ring by
blanking the according component in the individual channel maps
(Fig. \ref{fig:hi}). The velocity field of each component is well
ordered and shows the spider diagram typical of an inclined rotating
disk. We used tilted-ring fitting routines to derive the kinematic
parameters (inclination, position angle, dynamical center, systemic
velocity) and the HI rotation curve.  The position angle
($PA=35^o\pm1^o$) of the kinematic major axis of the outer HI ring is
aligned with the major axis of the S0 host disk ($PA \sim 37^o$;
Fig. \ref{fig:dss}). The position angle of the inner polar ring is
off-set by about 70$^o$. However, the inclination derived for the two
HI gas rings and the inner S0 host are similar ($i
\sim 65^o$) within the errors (see also Shane 1980, Mahon 1992).  For
the final rotation curve, we assumed that the rotation velocities in
the polar ring reflect those in the HI disk at the corresponding radii
(Fig. \ref{fig:dss}). The rotation curve is consistent with solid body
rotation out to a radius of $\sim 50'' (3.3~kpc)$ which includes the
position of the polar ring ($r\sim0.57'\approx 2.2\,kpc$).

{\it Outer HI ring:} The outer HI ring is situated at a radius of
$r\sim 2.4' (\approx 9.4\,kpc$) in a large-scale ($r \sim 2.6' \approx
10.1\,kpc$) diffuse disk.  The HI column density exceeds $\sim
1\times10^{21}\,\cms$ in a few locations within the outer ring at
scales of $\sim 10''$. The outer ring contains an atomic gas mass of
about $8.8\times10^8\,\solm$ or about 62\% of the total atomic gas
mass ($1.41\times10^9\,\solm$).  In the DSS2 red image a very faint
ring-like structure surrounding the main stellar disk (see also
Peletier \& Christodoulou 1993) coincides with the HI ring
(Fig. \ref{fig:dss}). The change in position angle and ellipticity at
$r\sim(80-90)''$ are indicative of a bar with a similar semi-major
axis length (Fig. \ref{fig:dss}). If we assume the standard relation
between the corotation resonance (CR) radius and the bar semi-major
axis $a$ ($r_{CR}=1.2a$), we find $r_{CR}\approx100''$ and a bar
pattern speed of $\Omega \sim 25 - 30\,\kms\,kpc^{-1}$. The position
of the outer Lindblad resonance (OLR) lies approximately at the radius
of the HI ring under these assumptions (Fig. \ref{fig:dss}).

\begin{figure}
\epsscale{0.9}
\plotone{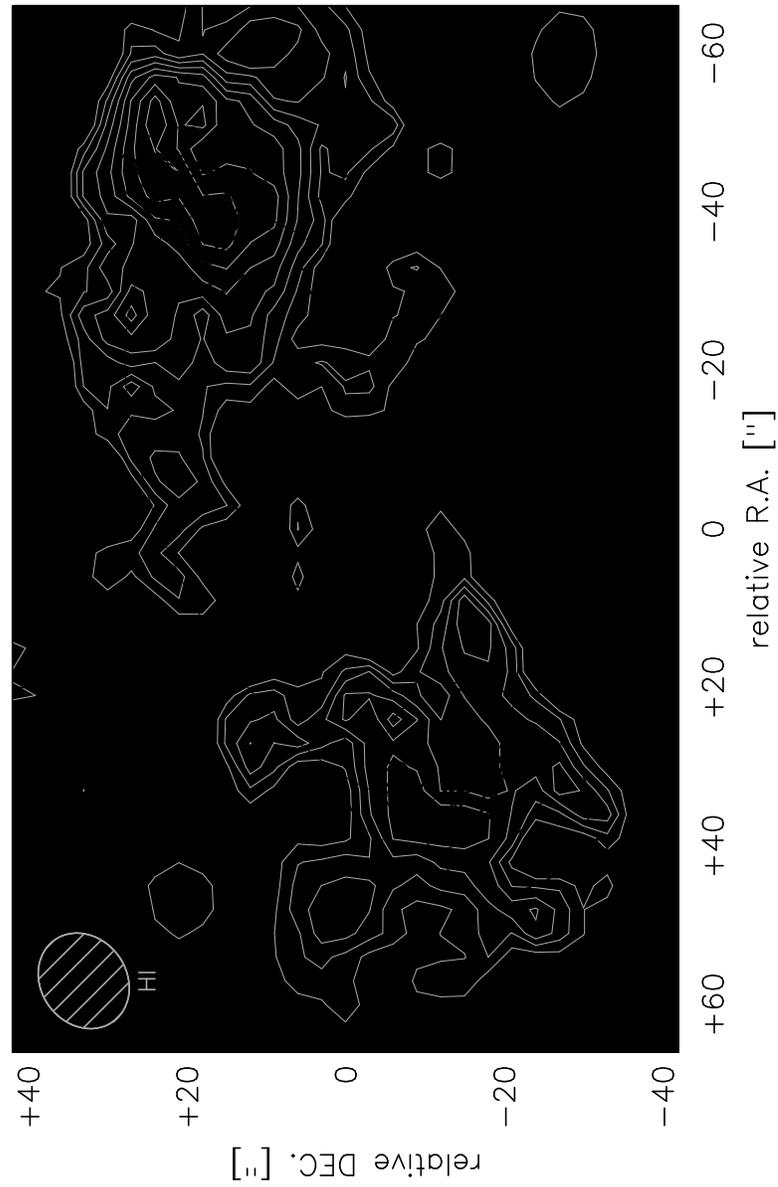}
\caption{
The molecular gas as seen in its CO line emission by OVRO (fat black
contours) is located close to the densest HI peaks (grey contours) and
the HII regions seen in the H$\alpha$ line emission (grayscale). The
IRAM 30m pointings are indicated by the broken circles
delineating the 3mm HPBW. The thin ellipse indicates the geometry of
the polar ring. The beam of the OVRO and HI data are shown.
\label{fig:co}}
\end{figure}

\begin{figure}
\epsscale{1.0}
\plotone{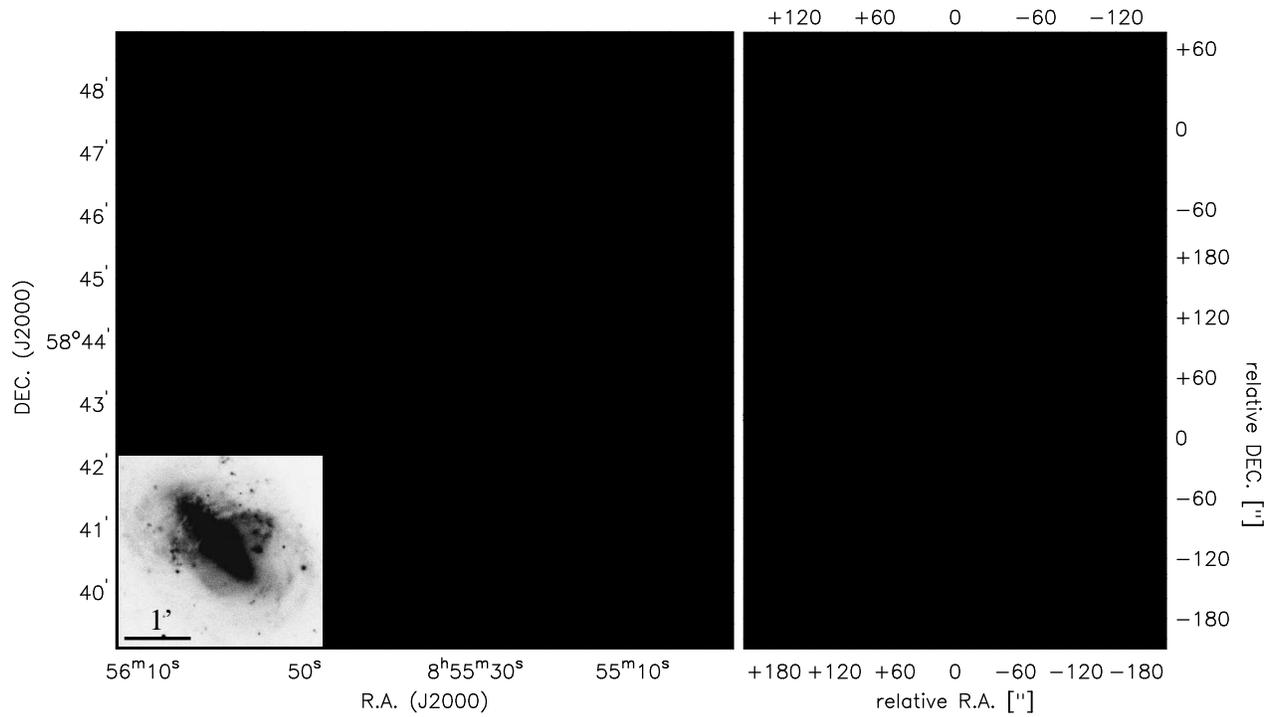}
\caption{
Intensity maps (grayscale) and velocity fields (contours) of the VLA
HI data. Increasing (solid line) (decreasing; broken line) velocities
are shown in steps of $20\,\kms$ relative to $v_{helio}=875\,\kms$
(thick line). {\it (a)} Total HI emission. The
inset shows an optical image (from NOAO/AURA/NSF) of NGC~2685. {\it
(b)} The polar ring component only. {\it (c)} The outer HI disk/ring.  
}\label{fig:hi}
\end{figure}

\begin{figure}
\epsscale{0.7}
\plotone{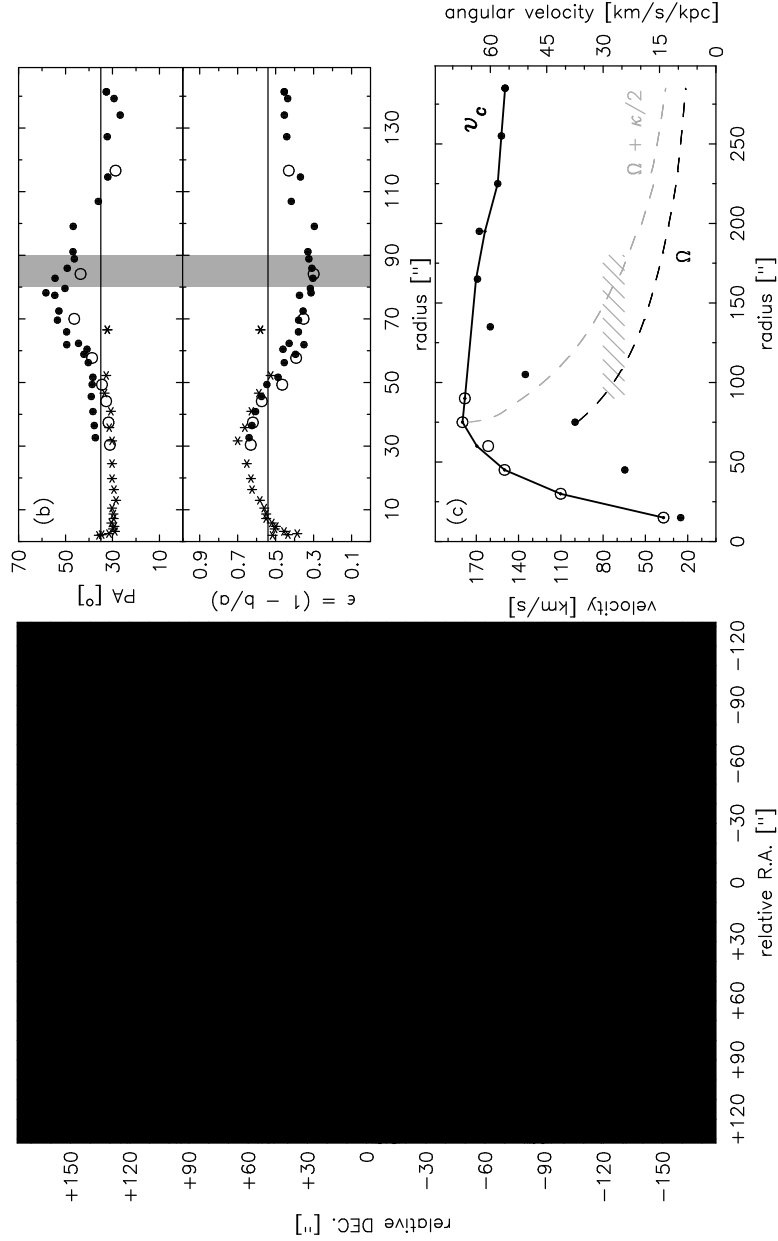}
\caption{
{\it (a)} Smoothed DSS2 red image (contours) overlaid on the HI outer
disk/ring component (see text). {\it (b)} The change in ellipticity
$\epsilon=(1-\frac{b}{a}$) and position angle of fitted ellipses to
the DSS2 red image (solid circles) indicate a bar semi-major axis of
$\sim 80''$ (shaded area). The values of Peletier
\& Christodoulou (1993) derived from a $K$ band image (stars) and a
deep $F$ band image (open circles) are shown as well. {\it (c)} The
rotation curve derived from the HI data (solid line) and the fits to
the natural weighted outer disk only (solid circles) and the polar
ring (open circles). The corresponding curves for
$\Omega$ (broken dark gray line) and $\Omega+\frac{\kappa}{2}$ (broken
light gray line) indicate that the OLR is at about 150'' (hatched
area) for a bar semi-major axis length of about 80''.
\label{fig:dss}}
\end{figure}

{\it Polar HI ring:} The H$\alpha$ image by Eskridge \& Pogge (1997)
shows about 20 HII regions which delineate the polar ring
(Fig. \ref{fig:co}).  The HI polar ring coincides with the H$\alpha$
and optical-continuum emission forming the polar ring
(Fig. \ref{fig:hi} and Fig. \ref{fig:co}). The polar HI ring contains
about $2.9\times10^8\,\solm$ or 20\% of the total atomic gas
mass. This is about 4\% of the dynamical mass at this radius
($M_{dyn}(r=34'')=7.4\times10^9\,M_{\odot}$; assuming spherical
symmetry). The average HI column density in the western part of the
polar ring is about $1.4\times10^{21}\,cm^{-2}$ (peak of
$2.1\times10^{21}\,cm^{-2}$). In the eastern part, the average HI
column density is just below $10^{21}\,cm^{-2}$ on scales of $\sim
10''$.

\subsection{The Molecular Gas}

\begin{figure}
\epsscale{0.7}
\plotone{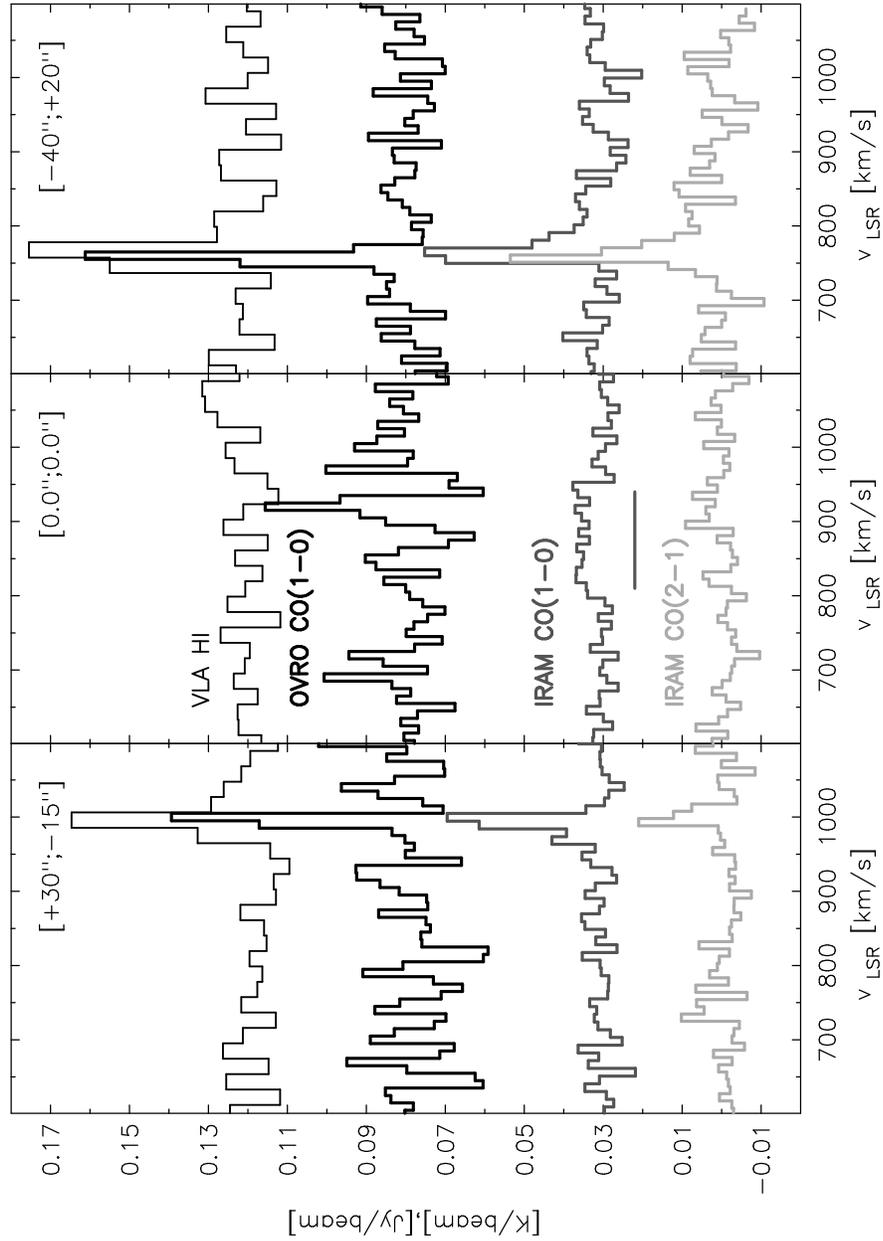}
\caption{
Comparison of the HI (thin black line) and CO spectra from the eastern
({\it left}) and western ({\it right}) part of the polar ring as well
as the nucleus ({\it middle}). Units are: HI ($[Jy/beam]\times9$, thin
black line), OVRO \CO\ ($[Jy/beam]$, thick black line), and IRAM \COe\
(thick dark grey line) and \COz\ (thick light grey line) in
$T_A^{\star} ([K/beam]$). The nuclear IRAM \COe\ spectrum shows a wide
($\sim 130\,\kms$) line (indicated by the bar) close to the 3$\sigma$
level.
\label{fig:spec}}
\end{figure}

\COe\ line emission is detected with the OVRO mm-interferometer in the 
eastern and western edge of the polar ring of NGC~2685
(Fig. \ref{fig:co}). The CO data are summarized in Table
1. No molecular line emission above 3$\sigma$ is seen from the nuclear
region, even when smoothed to a spectral resolution of 130\,$\kms$\
(1$\sigma\approx$6\,mJy/beam). The four GMAs in the polar ring are
spatially unresolved at our resolution of $\sim 6.6'' (\approx
430$\,pc). The molecular gas is located close to the densest HI peaks
in the polar ring (Fig. \ref{fig:co}) which are also next to the
brightest H$\alpha$ regions. The velocities and line widths of the
molecular gas agree well with those seen in the atomic gas of the
polar ring (Fig. \ref{fig:spec}), providing kinematic
confirmation that the CO emission is located in the polar ring and
not in the S0 host.

A comparison of the CO line flux detected in the OVRO data with the
IRAM 30m single dish line flux in the western (eastern) part of the
ring shows that the OVRO data recovers about 90 (40) \% of the single
dish flux (Table 1). Both \CO\ lines are detected in both sides of the
polar ring with the IRAM single dish telescope
(Fig. \ref{fig:spec}). The line FWHMs are fairly small ($20\,\kms$\
and $15\,\kms$\ for \COe\ and \COz). The properties of the GMAs (size
$\le 430\,pc$, velocity widths $\sim 15\,\kms$) are comparable to
those seen in other galaxies (e.g. in M~51, Aalto et al. 1999).  Using
the upper mass limit of $0.2\times10^6\solm$ from position W2 (Table
1), we derive an upper limit for the total undetected molecular gas
mass in the polar ring of about $6\times10^6\solm$. The central \COe\
IRAM spectrum clearly shows a $130\,\kms$ wide line at the $\sim
3\,\sigma$ level of $\sim 6\,mK$.  Using the relation
$S_O/S_I=(1+(\frac{\theta_S}{\theta_I})^2)/(1+(\frac{\theta_S}{\theta_O})^2)$
between the ratios of the fluxes ($S_{I,O}$ and the beam
($\theta_{I,O}$) and source sizes ($\theta_{S}$) (e.g. Dickel 1976)
for the IRAM (I) and OVRO (O) data, we find that the extent of the CO
emission must be of the order of $\sim 20''$ to explain the OVRO
non-detection.

\subsection{Comparison to previous CO detections}

Our CO line fluxes for the polar ring (region E1 and W1) are
considerably smaller than those previously reported by Watson, Guptill
\& Buchholz (1994) (their position 3 and 7). Given the good agreement
between our CO line widths ($\sim 20\,\kms$) and the HI line width
(compared to the CO line widths of $\sim 130\,\kms$ by Watson et
al. 1994) together with the consistency of the OVRO and IRAM data, we
conclude that the present data yield more accurate fluxes. The strong
\COe\ line emission apparent in the NRO 45m spectrum (Taniguchi et
al. 1990) is also inconsistent with our nuclear line flux.

\section{DISCUSSION AND IMPLICATIONS}

\subsection{Probable formation mechanism of the polar ring}

The (detected) molecular gas mass of the polar ring is only about 4\%
of the atomic mass present there ($log(M_{H_2}/M_{HI})\approx -1.40$), and
the ratio between the blue luminosity and the molecular gas mass
is also fairly low ($log(M_{H_2}/L_{B}\approx-2.7$) using
$L_{B}\sim4.5\times10^9\,\solar$ of Richter, Sackett \& Sparke
1994). These values are comparable to those found for low-mass ($\leq
10^{10}\,\solm$) very late-type galaxies (i.e. including dwarf
irregulars, Casoli et al. 1998) and S0 galaxies with counter-rotating
gas/stars (Bettoni et al. 2001).

The low far-infrared luminosity of $L_{FIR}\sim 4\times10^8\solar$
(using the IRAS fluxes and the standard relation for $L_{FIR}$;
Sanders \& Mirabel 1996) indicates no recent {\it massive} star
formation as observed in starburst galaxies in the NGC~2685
system. This is in agreement with the old age ($\sim 10\,Gyr$) of the
nuclear stellar population (Sil'chenko 1998). All this implies that
there was no considerable gas inflow into the central kiloparsec in
the recent past. This appears to contradict the models of the pole-on
disk merger (Bekki 1998) and polar encounter with a spiral galaxy
(Reshetnikov \& Sotnikova 1997), which show an accumulation of a few
$10^8\,\solm$ in the central kiloparsec. (Our derived molecular gas
mass in the inner kiloparsec is only about $4\times10^6\,\solm$.)

The host disk of NGC~2685 exhibits no obvious sign of an interaction
besides the more extended diffuse HI emission at the north-eastern
edge of the disk (Fig. \ref{fig:hi}). We can use equation (29) of Lake
\& Norman (1983) which relates the density of the accreted dwarf to
the mean density of the parent galaxy inside the radius of the
orbit. For a dynamical mass of $M_{dyn} (r=34'') = 7.4 \times 10^9
M_{\odot}$ inside the ring radius for the host galaxy, and a total
mass of $M_{dwarf}\sim 5\times M_{HI,ring}\approx1.5\times10^9\,\solm$
(assuming a typical scale factor of 5 between the HI mass
and the dynamical mass for dwarf galaxies; e.g. Swaters
1999), we find a radius for the accreted dwarf galaxy of about
2.2\,kpc. This radius and the HI mass of
$M_{HI}\sim3\times10^8\,\solm$ are typical values for late-type dwarf
galaxies (Mateo 1998, Swaters 1999).  Since the ring appears in its
colors neither very young ($< 1\,Gyr$) nor extremely ancient (Peletier
\& Christodoulou 1993), it seems likely that the polar ring was formed
a few Gyr ago during an accretion event (of a dwarf/low-mass galaxy)
similar to those proposed for S0 counter-rotators. The solar
metallicity derived for the ring HII regions (Eskridge \& Pogge 1997)
implies that either (1) the accreted dwarf galaxy had an elevated
intrinsic metallicity similar to those observed in a couple local
dwarf galaxies (e.g. Mateo 1998), (2) low-level continuous star
formation has enriched the polar ring ISM in the past few Gyr (see
e.g. Legrand et al. 2001) and/or (3) the ISM of the accreted dwarf
galaxy was mixed/enriched with more metal-rich material from the
primary galaxy.

The outer HI ring can be explained as gas accumulated at the OLR of
the 11\,kpc diameter bar. The good spatial correspondence of the
optical and HI rings suggests that there is enhanced star
formation. We conclude that the suggested scenario by Peletier \&
Christodoulou (1993) seems unlikely, where the outer HI ring has been
formed during the secondary accretion event which formed the inner
polar ring as well.

\subsection{Stability of the polar ring}

The OVRO map of the molecular gas provides the first unambiguous
(kinematically confirmed) detection of molecular gas in a polar
ring. This molecular gas is associated with about four GMAs which are
located close to on-going star formation and peaks in the atomic gas
which exceed $10^{21}\,\cms$. Note that the molecular and the atomic
gas appear concentrated where the polar ring intersects the disk of
the parent galaxy. 

The polar ring exhibits at least two stellar populations. The HII
regions present in the polar ring have an average age of 5\,Myr (if
instantaneous star formation is assumed) and about solar metallicity
(Eskridge \& Pogge 1997). However, Peletier \& Christodoulou (1993)
deduced an age of a few Gyr from the red colors of the polar ring.
The contribution of the atomic gas in the polar ring to the total
dynamical mass (enclosed out to its radius) is about 4\%. This might
already be enough to allow for self-gravitation of the polar ring, and
explain its persistence (Sparke 1986). Alternatively, Mahon (1992) has
found that the HI distribution and kinematics of NGC~2685 can be
stable using models of prograde anomalous orbits in a triaxial
gravitational potential.

Thus we conclude that: (a) the polar ring has been stable for a
substantial time (few Gyr) and (b) the recent star formation in the
polar ring has been triggered by another mechanism than the actual polar
ring formation process.

\acknowledgments
Special thanks to C. Thum for taking the IRAM data and to F. Schweizer
for his critical comments on an early draft. We like to thank
P. Eskridge for providing us with the H$\alpha$ image. ES acknowledges
support by National Sciene Foundation grant AST 96-13717.

\clearpage

\begin{table}
\begin{center}
\caption{Summary of CO (1-0) Data\label{tab:hi}}
\begin{tabular}{l|rrrrrr|rrrrrrrrrrr}
\tableline\tableline
Instrument &\multicolumn{6}{c}{IRAM} &\multicolumn{7}{c}{OVRO} \\
Component & N & W1 & W2 & E1 & E2 & E3 & N & W$_{N}$ & W$_{S}$ & W$_{tot}$ & E$_{N}$ & E$_{S}$ & E$_{tot}$ \\
\tableline
I$_{CO}$ [$K\,km\,s^{-1}$]       & 1.01 & 1.47 & $\le$ 0.46 & 1.13 & 1.10 & 1.18 & $\le$ 1.66 & 5.13 & 6.79 & 13.8 & 1.17 & 3.43 & 4.74 \\
S$_{CO}$ [$Jy\,km\,s^{-1}$]      & 4.82 & 7.02 & $\le$ 2.20 & 5.38 & 5.26 & 5.62 & $\le$ 0.78 & 2.41 & 3.20 & 6.50 & 0.55 & 1.61 & 2.23 \\
N$_{H_2}$ [10$^{21}$ cm$^{-2}$] & 0.20 & 0.29 & $\le$ 0.01 &      & 0.23 &      & $\le$ 0.33 & 1.03 & 1.36 & 2.76 & 0.23 & 0.69 & 0.95 \\
M$_{H_2}$ [10$^6$ M$_{\odot}$]  & 4.4  & 6.5  & $\le$ 0.2  &      & 5.0  &      & $\le$ 0.7  & 2.2  & 3.0  & 6.0  & 0.5  & 1.5  & 2.1  \\
\tableline\tableline
\end{tabular}
\tablecomments{
Properties of the molecular gas in NGC~2685 as derived from the single
dish and interferometric data. (The uncertainties in the flux
measurements are about 10\% (15\%) for the IRAM (OVRO) data.) The
total mapped molecular gas mass in the ring is $8.0\times10^6\,\solm$
compared to $\sim 10^7\,\solm$ detected by the IRAM single dish. The
OVRO upper limit for the nucleus is derived via $S_{CO}=3\sigma\times
\Delta v$ with $\Delta v=130\,\kms$ from the 30m spectrum. Assuming
that the average polar ring GMA has a small line width of $\sim
20\,\kms$, our upper mass limit is $0.4\times10^6\,\solm$ from the
OVRO data. We use a standard conversion factor of $X =
2\times10^{20}\,cm^{-2}/(K\,\kms)$ (Strong et al. 1987).  }
\end{center}
\end{table}

\clearpage

\clearpage

\clearpage

\clearpage 

\end{document}